\documentstyle[sprocl, epsfig]{article}
\def\be{\begin{equation}}
\def\ee{\end{equation}}
\def\bea{\begin{eqnarray}}
\def\eea{\end{eqnarray}}
\def\beq{\begin{equation}}
\def\eeq{\end{equation}}
\def\bq{\begin{quote}}
\def\eq{\end{quote}}
\def\gappeq{\mathrel{\rlap {\raise.5ex\hbox{$>$}} {\lower.5ex\hbox{$\sim$}}}}
\def\lappeq{\mathrel{\rlap{\raise.5ex\hbox{$<$}} {\lower.5ex\hbox{$\sim$}}}}
\def\PR{{\it Phys.~Rev.~}}

\def\NP{{\it Nucl.~Phys.~}}

\def\PL{{\it Phys.~Lett.~}}

\def\SJNP{{\it Sov.~J.~Nucl.~Phys.~}}
\def\SPJETP{{\it Sov.~Phys.~J.E.T.P.~}}
\def\ZP{{\it Zeit.~Phys.~}}

\def\JHEP{{\it J.~High~En.~Phys.~}}
\def\vol#1{{\bf #1}}
\def\vyp#1#2#3{\vol{#1} (#2) #3}

\def\epm#1#2{\hbox{${\lower1pt\hbox{$\scriptstyle +#1$}}
\atop {\raise1pt\hbox{$\scriptstyle -#2$}}$}}

\def\gsim{\mathrel{\rlap{\lower4pt\hbox{\hskip1pt$\sim$}}
    \raise1pt\hbox{$>$}}}         

\def\ie{{\it i.e.}}

\def\etal{{\it et al.}}

\def\toinf#1{\mathrel{\mathop{\sim}\limits_{\scriptscriptstyle {#1\rightarrow\infty }}}}

\def\frac#1#2{{{#1}\over {#2}}}
\def\half{\hbox{${1\over 2}$}}

\def\smallfrac#1#2{\hbox{${{#1}\over {#2}}$}}

\def\as{\alpha_s}

\def\GeV{{\rm GeV}}
\def\MS{\hbox{$\overline{\rm MS}$}}

\def\bq{\bar{q}}
\catcode`@=11 
\def\slash#1{\mathord{\mathpalette\c@ncel#1}}
 \def\c@ncel#1#2{\ooalign{$\hfil#1\mkern1mu/\hfil$\crcr$#1#2$}}
\def\lsim{\mathrel{\mathpalette\@versim<}}
\def\gsim{\mathrel{\mathpalette\@versim>}}
 \def\@versim#1#2{\lower0.2ex\vbox{\baselineskip\z@skip\lineskip\z@skip
       \lineskiplimit\z@\ialign{$\m@th#1\hfil##$\crcr#2\crcr\sim\crcr}}}
\catcode`@=12 

\begin{document}
\begin{titlepage}
\setcounter{page}{0}
\topmargin 2mm
\begin{flushright}
{\tt hep-ph/0001157}\\
{CERN-TH/2000-010}\\ {Edinburgh 2000/01}
\\ {RM3-TH/00-01}
\end{flushright}
\vskip 12pt
\begin{center}

{\bf SINGLET PARTON EVOLUTION AT SMALL $x$:\\ a Theoretical Update} 
\vskip 12pt
{ Guido Altarelli$^{1,\,2}$,  Richard D. Ball$^{4}$\footnote[1]{Royal Society University
Research Fellow}  and   Stefano Forte$^{3}$\footnote[2]{On leave from INFN, Sezione di Torino, Italy}}
\vskip 6pt

{\em  {}$^1$Theory Division, CERN\\ CH--1211 Geneva 23, Switzerland \\
\vskip 4pt{}$^2$Universit\`a di Roma Tre\\ {}$^3$INFN, Sezione di Roma Tre\\ Via della Vasca Navale
84, I-00146 Rome, Italy\\
\vskip 4pt{}$^4$Department of Physics and Astronomy, University of Edinburgh,\\ Mayfield Road, Edinburgh EH9
3JZ, Scotland\\} 
\addtocounter{footnote}{-2}
\vskip 30pt
\end{center}
{\narrower\narrower
\centerline{\bf Abstract}
\medskip\noindent

This is an extended and pedagogically oriented version of our recent
work, in which
we proposed an improvement of the splitting functions at small
$x$ which overcomes the apparent problems encountered by the BFKL approach.
\\
}

\smallskip
\vskip 20pt
\begin{center}
{Presented by G. Altarelli at the XVII Autumn School\\
"QCD: Perturbative or Nonperturbative?"\\ 
Instituto Superior Tecnico, Lisbon, Portugal,
October 1999}                                 
\end{center}
\vfill
\begin{flushleft} CERN-TH/2000-010\\ January 2000 \end{flushleft} 
\end{titlepage}


\title{SINGLET PARTON EVOLUTION AT SMALL $x$:\\ a Theoretical Update}

\author{Guido Altarelli,$^{1,\,2}$  Richard D. Ball$^{4,\;}$\footnote[1]{Royal Society University
Research Fellow},  
 Stefano Forte$^{3,\;}$\footnote[2]{On leave from INFN, Sezione di Torino, Italy}}
\vskip4pt
\address{$^1$Theory Division, CERN\\ CH--1211 Geneva 23, Switzerland
\\
\vskip4pt
 {}$^2$Universit\`a di Roma Tre \\ {}$^3$INFN, Sezione di Roma Tre\\ Via della Vasca Navale
84, I-00146 Rome, Italy\\ \vskip4pt{}$^4$Department of Physics and Astronomy, University of Edinburgh,\\ Mayfield Road, Edinburgh EH9
3JZ, Scotland}

\maketitle\abstracts{ This is an extended and pedagogically oriented version of our recent work in ref.~[1] where 
we proposed an improvement of the splitting functions at small
$x$ which overcomes the apparent problems encountered by the BFKL approach.}
\section{ Introduction}
\noindent

The theory of scaling violations in deep inelastic scattering  is one of the most solid consequences of
asymptotic freedom and provides a set of fundamental tests of QCD. At large $Q^2$ and not too small but fixed $x$ the
QCD evolution equations for parton densities~\cite{glap} provide the basic framework for the description of scaling
violations. The complete splitting functions have been computed in perturbation theory at order
$\alpha_s$ (LO  approximation) and $\alpha_s^2$ (NLO)~\cite{nlo}. For the first few moments the anomalous dimensions 
at order $\alpha_s^3$  are also known~\cite{nnlo}. 

At sufficiently small $x$ the approximation of the splitting functions based on  the first few terms of the expansion 
in powers of $\alpha_s$ is not in general a good approximation.  If not for other reasons, as soon as $x$ is small
enough that 
$\alpha_s \xi\sim 1$, with $\xi=\log{1/x}$, all terms of order
$\alpha_s (\alpha_s \xi)^n$ and $\alpha_s^2 (\alpha_s \xi)^n$ which are present~\cite{kis} in the splitting functions
must be considered in order to achieve an accuracy up to  order
$\alpha_s^3$. In terms of the anomalous dimension $\gamma(N,\alpha_s)$, defined as the $N$--th  Mellin moment of the
singlet splitting function (actually the eigenvector with largest eigenvalue), these terms correspond to sequences of
the form $(\alpha_s/N)^n$ or $\alpha_s(\alpha_s/N)^n$.   In most of the kinematic region of HERA~\cite{klein} the
condition $\alpha_s
\xi\sim 1$ is indeed true. Considering that, by kinematics, $\as \xi\leq \as \log{s/Q^2}$, and $s\approx
10^5~GeV^2$, $\as(m_Z^2)\approx 0.119$, we see that at $Q^2~=~3, 10, 10^2, 10^3~GeV^2$ $\as \xi$ can be as large as
$\as
\xi\leq 4.3,~ 3.0, ~1.2, ~0.6$, respectively. Hence,  in principle one could expect to see in the data indications of
important corrections to the approximation~\cite{DGPTWZ,das}  of splitting functions computed  only  up to order
$\alpha_s^2$ and the corresponding small $x$ behaviour. In reality this appears not to be the case: the data can be
fitted quite well by the evolution equations in the NLO approximation~\cite{das,dero}. 
An idea of the quality of the fit can be obtained from fig.~1 where a comparison of the data with the QCD NLO scaling
violations is displayed. Of course it may be that some corrections exist but they are hidden in a redefinition of the
gluon, which is the dominant parton density at small $x$. While the data do not support the presence of large corrections
in the HERA kinematic region~\cite{bfklfits} the evaluation of the  higher order corrections at small $x$ to the singlet
splitting function from the BFKL theory~\cite{bfkl,jar,ktfac} appears to fail.  The results of the recent
calculation~\cite{fl,cc,dd} of the NLO term
$\chi_1$ of the BFKL function
$\chi=\alpha_s\chi_0+\alpha_s^2\chi_1...$ show that the expansion is very badly behaved, with the non leading term
completely overthrowing the main features of the leading term. Taken at face value, these results appear  to hint at
very large corrections to the singlet splitting function at small $x$ in the region explored by HERA~\cite{flph}.
\begin{figure}
\begin{center}
\mbox{
 \epsfig{file=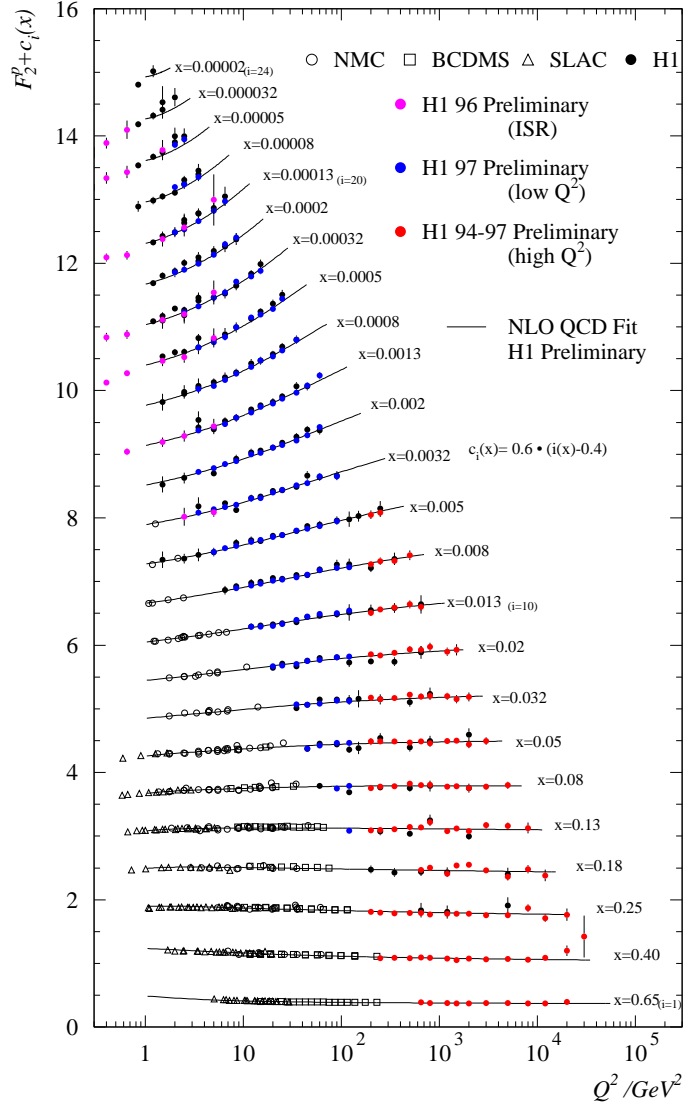,width=0.9\textwidth}}
\vskip.3truecm
\caption{\baselineskip 10pt 
Next-to-leading order QCD fit to HERA and pre-HERA data on the
proton structure function $F_2(x,Q^2)$~[9]. }{ }
\end{center}
\end{figure}

Here we review this problem and propose a  procedure to construct a meaningful improvement of the singlet
splitting function at small $x$, using the information from the BFKL function. After recalling some basic notions we
start by defining an alternative expansion for the BFKL function
$\chi(M)$ which, unlike the usual expansion, is well behaved and stable when going from LO to NLO, at least for
values far from $M=1$. This is obtained by adding suitable sequences of terms of the form $(\alpha_s/M)^n$ or
$\alpha_s(\alpha_s/M)^n$ to
$\as\chi_0$ or $\as^2\chi_1$ respectively. The coefficients are determined by the known form of the singlet anomalous
dimension at one and two loops.  This amounts to a resummation~\cite{salam} of $(\alpha_s\log{Q^2/\mu^2})^n$ terms in
the inverse $M$-Mellin transform space. This way of improving $\chi$ is completely analogous to the usual way   of
improving~$\gamma$~\cite{summ}. One important point, which is naturally reproduced with good accuracy by the above
procedure, is the observation that the value of $\chi(M)$ at $M=0$ is fixed by momentum conservation to be
$\chi(0)=1$. This observation plays a crucial role in formulating the novel expansion  and explains why the normal
BFKL expansion is so unstable near $M=0$, with $\chi_0\sim 1/M$, $\chi_1\sim -1/M^2$ and so on.  This rather
model--independent step is already sufficient to show that no catastrophic deviations from the NLO approximation of
the evolution equations are to be expected. The next step is to use this novel expansion of  $\chi$ to determine
small $x$ resummation corrections  to add to the LO and NLO anomalous dimensions $\gamma$. Defining $\lambda$ as the
minimum value of $\chi$,
$\chi(M_{min})=\lambda$, and using the results of ref.~\cite{sxap}, a meaningful expansion for the improved anomalous
dimension is written down in terms of $\chi_0$,
$\chi_1$, and $\lambda$. The large negative correction to $\lambda_0/\as=\chi_0(1/2)$ induced by $\as\chi_1$,  that
is formally of order $\as$ but actually is of order  one for the relevant values of $\alpha_s$, suggests that
$\lambda$ should be reinterpreted as a nonperturbative parameter.  We conclude by showing that the very good
agreement of the data  with the NLO evolution equation can be obtained by choosing a  small value of $\lambda$, 
compatible with zero.

\section{ $Q^2$ Evolution at Small $x$}
\noindent

The behaviour of structure functions at small $x$ is dominated by the singlet parton
component.  Thus we consider the singlet parton density
\beq G(\xi,t)=x[g(x,Q^2)+k_q\otimes q(x,Q^2)],\label{Gdef}
\eeq where $\xi=\log{1/x}$, $t=\log{Q^2/\mu^2}$, 
$g(x,Q^2)$ and $q(x,Q^2)$ are the gluon and singlet quark parton densities, respectively, and $k_q$ is such that, for
each moment
\beq G(N,t)=\int^{\infty}_{0}\! d\xi\, e^{-N\xi}~G(\xi,t),\label{Nmom}
\eeq the associated anomalous dimension $\gamma(N,\as(t))$ corresponds to the largest eigenvalue in the singlet
sector. At large $t$ and fixed $\xi$ the evolution equation in $N$-moment space is then
\beq
\frac{d}{dt}G(N,t)=\gamma(N,\as(t))~G(N,t),\label{tevol}
\eeq where $\as(t)$ is  the running coupling. The anomalous dimension is completely known  at one and two loop level:
\beq
\gamma(N,\as)=\as\gamma_0(N)+\as^2\gamma_1(N)+
\dots\>.\label{gamexp}
\eeq As $\gamma(N,\as)$ is, for each $N$, the largest eigenvalue  in the singlet sector, momentum conservation order
by order  in $\as$ implies that
\beq
\gamma(1,\as)=\gamma_{0}(1)=\gamma_{1}(1)=...=0.\label{mcons}
\eeq

The solution of eq.(\ref{tevol}) is given by
\beq
G(N,t)=G(N,0)~\exp{\int^{\as(t)}_{\as}~d\as'~\frac{\gamma(N,\as')}{\beta(\as')}}\label{sol}
\eeq
where $\as\equiv\as(0)$ and the QCD beta function $\beta(\as)$ is defined by 
\beq
\frac{d\as(t)}{dt}=\beta(\as(t));~~~~~\beta(\as)=-b\as^2(1+b'\as+.....);~~~~~
b=\frac{\beta_0}{4\pi}=\frac{11 n_c-2 n_f}{12\pi}\label{beta}
\eeq
At one loop accuracy $\gamma(N,\as)=\as\gamma_0(N)$ and the solution reduces to:
\beq
G(N,t)=G(N,0)~e^{(\gamma_0\eta/b)}=G(N,0)~[\frac{\as}{\as(t)}]^{\gamma_0/b};~~~~\eta=\log{\frac{\as}{\as(t)}}\label{sol1l}
\eeq
Of formal interest is the solution in the limit of fixed coupling. In general, in this limit,  we have
$G(N,t)=G(N,0)~\exp{[\gamma(N,\as)t]}$ and, at one loop, $G(N,t)=G(N,0)~\exp{(\gamma_0\as t)}$. Comparing with the
running coupling case at one loop, we see that $\as t$ at fixed coupling is replaced by $\eta/b$ at running coupling. In
fact by expanding we have
\beq
\eta=\log{\frac{\as}{\as(t)}}=\log{(1+b\as t)}=b\as t+......\label{etadef}
\eeq
It is useful to keep in mind this translation rule for one loop results. 

We now consider the small $x$ behaviour of the one loop solution. Small $x$ means large $\xi$, hence small $N$. At
small $N$,
$\gamma_0(N)=\frac{n_c}{\pi N}+\ldots$. The $1/N$ behaviour correspond to the $1/x$ behaviour of the one
loop singlet splitting function $P_0(x)$, the Mellin transform of $\gamma_0$: $\gamma(N) =\int^1_0\!dx\,x^N\!P(x)$. 
In general $P(x)\sim 1/x(\log(1/x))^n$ corresponds to $\gamma(N)\sim n!/N^{n+1}$. We can add a constant term to
$\gamma_0(N)$ in such a way that the momentum conservation condition at $N=1$ is respected:
\beq
\gamma_0(N)\approx\frac{n_c}{\pi}(\frac{1}{N}-1)+.....\label{ccc}
\eeq
This addition provides a good approximation in a wide region
extending from $N\sim 0$ up to $N\sim 1$. From eq.(\ref{sol1l}) one
obtains the leading behaviour: 
\beq
G(N,t)=G(N,0)~\exp\left[\frac{n_c}{\pi}\frac{\eta}{b}(\frac{1}{N}-1) \right]\label{lb} 
\eeq
By taking the inverse Mellin transform we
obtain:
\beq
G(\xi,t)=\int_{c-i\infty}^{c+i\infty}\!\frac{dN}{2\pi i}\!\exp{(N\xi+\frac{n_c\eta}{\pi b}(\frac{1}{N}-1) )}
G(N,0)\label{MT}
\eeq
where the integral is along the imaginary axis at $RealN=c$. The leading behaviour
at small $x$ is obtained by the saddle point method (see Appendix). Let $N_0$ be the point where the derivative of
the exponent vanishes:
\beq
\frac{d}{dN}\left(N\xi+\frac{n_c\eta}{\pi b }
\left(\frac{1}{N}-1\right)\right)_{N=N_0}=0,~~~~N_0=\sqrt{\frac{n_c\eta}{\pi b \xi}}\label{Nzero}
\eeq
At $N_0$ the value of the exponent is:
\beq
\left(N\xi+\frac{n_c\eta}{\pi b }\left(\frac{1}{N}-1\right)\right)_{N=N_0}=\sqrt{\frac{4n_c\eta \xi}{\pi b}}-\frac{n_c\eta}{\pi b}\label{fN}
\eeq
 For $G(N,0)$ regular at $N\geq 0$, the resulting small $x$ behaviour is given by:
\beq
G(\xi,t)=G(N_0,0)\eta^{\smallfrac{1}{4}}\xi^{-\smallfrac{3}{4}}\exp{\sqrt{\frac{4n_c\eta\xi}{\pi b}}}
\exp{(-\frac{n_c\eta}{\pi b})}\label{res}
\eeq
This is the well known "double scaling" behaviour, first predicted in
ref.\cite{DGPTWZ} and developed in
refs.\cite{das}. The exponential factor
\beq
\exp{\sqrt{\frac{4n_c\eta\xi}{\pi b}}}\approx\exp{\sqrt{\frac{4n_c}{\pi
b}\log{\frac{1}{x}}\log{\frac{\log{Q^2/\Lambda^2}}{\log{\mu^2/\Lambda^2}}}}} \label{exp}
\eeq
produces a peak at small $x$ which increases with $Q^2$. The $x$
behaviour is weaker than any power of $x$ but stronger
than any power of a logarithm. This one loop prediction is not much modified by NLO corrections because the computed
two loop anomalous dimension $\gamma_1(N)$ also behaves like $1/N$. And in fact the complete NLO QCD evolution
reproduces the double scaling exponential rise quite accurately. As already mentioned the HERA data are quite well
fitted for
$Q^2\geq~1-4~\GeV^2$ by QCD $Q^2$ evolution computed at NLO accuracy. This agreement of the data with the usual
evolution equations is however rather unexpected. In fact the $Q^2$ evolution takes into account all terms of
order $[\as\log{1/x}\log{Q^2}]^n$ and $[\as\log{Q^2}]^n$ and are valid for sufficiently large $Q^2$ and $x$ small but
fixed. However possible terms of order $[\as\log{1/x}]^n$ are not included in the splitting function, so that the
results are not necessarily reliable for sufficiently small $x$ at fixed $Q^2$. The powerful approach based on the
BFKL equation provides a suitable tool towards an estimate of these terms and will now be discussed.

\section{The BFKL Function}
\noindent

The starting point of the BFKL theory is that, at large $\xi$ and fixed $Q^2$ (in
particular at fixed $\as$), the  following $x$ evolution equation for
$M$ moments was proven valid:
\beq
\frac{d}{d\xi}G(\xi,M)=\chi(M,\as)~G(\xi,M),\label{xevol}
\eeq where
\beq G(\xi,M)=\int^{\infty}_{-\infty}\! dt\, e^{-Mt}~G(\xi,t),\label{Mmom}
\eeq
For convergence, we consider $0<M<1$ as the physical region  and assume that $G(\xi,t)$ vanishes as $Q^2$ in the
photoproduction limit $Q^2\rightarrow\!0$ and approaches a constant modulo logs in the limit of large $Q^2$. 
In eq.(\ref{xevol}) $\chi(M,\as)$ is the BFKL function which is now known at NLO accuracy:
\beq
\chi(M,\as)=\as\chi_0(M)+\as^2\chi_1(M)+\dots\>.\label{chiexp}
\eeq
The BFKL function has been computed in perturbation theory at one and two loops starting from the large $s$ behaviour
of the amplitude for gluon-gluon scattering. 

We consider $T$, the absorptive part of the $A+B\rightarrow A+B$ forward
scattering amplitude from gluon exchange, as depicted in fig. 2. $A$ and $B$ are generic real or virtual
particles. In the case of interest, $A$ will be the virtual photon and $B$ the proton. From the optical theorem $T$ is
proportional to the total $ A+B$ cross section $\sigma$. We are interested in the gluon exchange component which is
dominant in the singlet sector. So in the deep inelastic scattering case the virtual photon $A$ is attached to a quark loop
which is included in $\phi_A$. Going down vertically then there is the g-g amplitude $\Gamma_N$ and finally the proton blob
$\phi_B$. The cross section $\sigma$ can be written as:
\beq
\sigma\sim\int\!\frac{d^2k_1}{k_1^2}\!\phi_A(\vec k_1)\int\!\frac{d^2k_2}{k_2^2}\!\phi_B(\vec k_2)
\int_{a-i\infty}^{a+i\infty}\!\frac{dN}{2\pi i}~(\frac{s}{k_1k_2})^N~\Gamma_N(\vec k_1,\vec k_2)\label{gg}
\eeq
Here $k_{1,2}$ are transverse momentum vectors of the s-channel gluons with virtuality $-(\vec k)^2$, $\phi_{A,B}$ are
suitable hadronic structure functions, $\Gamma_N$ is the $g$-$g$ kernel and $s=2p_Ap_B$ is the squared invariant mass of
the colliding particles. Note that a symmetric scale choice was
adopted in the $s$ factor. The $g$-$g$  kernel in the
$\MS$ scheme obeys the generalised BFKL equation:
\beq
N\Gamma_N(\vec k_1,\vec k_2)=\delta^{(2)}(\vec k_1-\vec k_2)+\int\!d^2k~ K(\vec k_1,\vec k)~ \Gamma_N(\vec
k,\vec k_2)\label{gg1}
\eeq

The BFKL equation can be cast in the form of the $x$ evolution 
equation~(\ref{xevol})  inverting the $N$-Mellin transform
eq.~(\ref{Nmom}) and taking an $M$-Mellin transform
eq.~(\ref{Mmom}). The inversion of the $N$-Mellin turns multiplication
by $N$ into differentiation with respect to $\ln (1/x)$, and the
$M$-Mellin turns the convolution with respect to $k^2$ into an
ordinary product. It then follows that the
 $\chi(M,\as)$ function is related to $K$ by $M$-Mellin transformation:
\beq
\chi(M,\as)=\int\!d^2k~ K(\vec k_1,\vec k)~ (\frac{k^2}{k_1^2})^{M-1}\label{gg2}
\eeq
At NLO and beyond, $\chi(M,\as)$ can acquire a  further dependence on
$k_1^2$ due to the running of the coupling. From the symmetry
under $\vec k_1\rightarrow \vec k_2$ exchange of $\Gamma_N$, the symmetry of $K$ also follows and in turn, at leading
order, this implies that $\chi_0(M)=\chi_0(1-M)$. We have slightly
cheated in this derivation in that the variable $N$ in eq.~(\ref{gg1})
is defined by the moment integration of eq.~(\ref{gg}), i.e. with the
symmetric factor $(\frac{s}{k_1k_2})^N$. The $N$ moments
eq.~(\ref{Nmom}) instead are taken with the factor $x^{N}=
(\frac{s}{Q^2})^{-N}$ where $Q^2$ is the virtuality of one of the two
particles, say $Q^2=k_1^2$. This difference is irrelevant at LO, but
at NLO it must be taken into account and modifies~\cite{fl} 
the relation between
$\chi$ and $K$ eq.~(\ref{gg2}).

\begin{figure}[t!]
\begin{center}
    \mbox{
      \epsfig{file=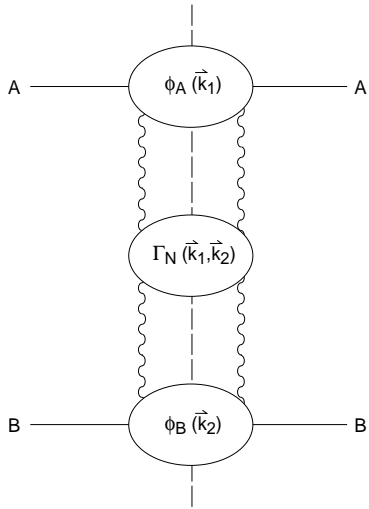,width=0.4\textwidth}
      }\caption{\baselineskip 10pt 
The process $A+B\rightarrow A+B$. }
\end{center}
\end{figure}

At one loop accuracy $\chi_0(M)$ is scheme and scale independent and
has the following simple form (see fig.4 below):
\beq
\chi_0(M)=\frac{n_c}{\pi}\int_0^1~\frac{dz}{1-z}~[z^{M-1}+z^{-M}-2]\label{chi0}
\eeq
(an alternative expression is $\chi_0(M)=n_c/\pi [ 2\psi(1)-\psi(M)-\psi(1-M)]$ with
$\psi(M)=\Gamma'(M)/\Gamma(M)$, where $\Gamma$ is the Euler gamma function). 
By expanding the denominator and performing the integration, one
finds the expression:
\beq
\chi_0(M)=\frac{n_c}{\pi}\sum_{n=0}^\infty[\frac{1}{M+n}+\frac{1}{n+1-M}-\frac{2}{n+1}]=\frac{1}{M}+2\zeta(3)
M^2 +2\zeta(5)M^4+.....\label{chi01}
\eeq
Also note that 
\beq
\chi_0(\smallfrac{1}{2})=\frac{n_c}{\pi}4\ln{2}\label{chi02}
\eeq

In eq.~(\ref{xevol}) the coupling $\as$ is fixed.  The inclusion of
running effects in the BFKL 
theory is a
delicate point. If we are only interested to next-to-leading log $1/x$
accuracy, the running of the coupling in the evolution equation~(\ref{tevol})
can be expanded out to leading order in $t$:
$\alpha_s(t)=\alpha_s(\mu^2)(1- b \alpha_s(\mu^2) t)$. Upon $M$-mellin
transformation, this corresponds to a differential operator:
$\alpha_s(M)=\alpha_s
(1+ b \alpha_s {d\over d M})$, which can be used in the $x$ evolution
equation~(\ref{xevol}). This affects the relation between the $t$ and
$x$ evolution equations, and in particular it modifies the relations
between $\chi$ and the anomalous dimension $\gamma$ which we will
derive in the next section (duality relation). However, for all
practical purposes, this modification can be simply viewed as the
result of an additional contribution to $\chi$: once this contribution
is incorporated in $\chi$, the anomalous dimension $\gamma$ is given
by the duality relation  given in the next session.  
Henceforth, we assume that $\chi$ includes such contribution, at the
appropriate order in the  small $x$
expansion. 
To next-to-leading order in
$\as$ (i.e. to NLLx), this corresponds to a contribution to 
$\chi_1$ proportional to the first coefficient 
$\beta_0=\smallfrac{11}{3}n_c-\smallfrac{2}{3}n_f$ of the 
$\beta$-function. 
Since the extra term depends on  the definition of the gluon density, it is also
necessary to  specify the choice of factorization scheme: 
here we choose the \MS\ scheme, so that the $\chi_1$ that
we will  consider in the sequel is given by~\cite{sxap}
\beq
\chi_1(M) = \smallfrac{1}{4\pi^2} n_c^2 \tilde\delta(M)  +\smallfrac{1}{8\pi^2}\beta_0 n_c
((2\psi'(1)-\psi'(M)-\psi'(1-M)) +\smallfrac{1}{4n_c^2}\chi_0(M)^2,\label{chionedef}
\eeq where the function $\tilde\delta$ is defined in the first of  ref.~\cite{fl}. 
 Note that 
 beyond the leading order the
symmetry of the BFKL function under $M \rightarrow 1-M$ is destroyed by the effects of running and by the fact that
the scales $k_1$ and $k_2$ are very different in the actual case of interest. As a result the actual BFKL function
which is relevant for deep inelastic leptoproduction is not perturbative near $M=1$, the point where the
soft scale is approached.

To start exploring the physical significance of the $x$ evolution equation, we work in the fixed coupling
approximation. The solution of eq.(\ref{xevol}) is given by:
\beq
G(\xi,M)=G(0,M)\!\exp{\left[\chi(M,\as)\xi\right]}\label{xevols}
\eeq
By taking the inverse Mellin transform, we obtain:
\beq
G(\xi,t)=\int_{c-i\infty}^{c+i\infty}\!\frac{dM}{2\pi i}\!\exp{[Mt+\chi(M,\as)\xi]} G(0,M)\label{MTx}
\eeq
For $t$ and $\xi$ large we consider the saddle point condition (for simplicity, we restrict our discussion to the LO):
\beq
\frac{d}{dM}\bigg ( Mt+\as\chi_0(M)\xi \bigg)_{M=M_0}=0,
~~~~\rm{or}~~~~\frac{t}{\xi}=-\as\chi'_0(M_0)\label{Mzero}
\eeq
 
With $t$ and $\xi$ both large we can study three different interesting limits: 1) $t/\xi$ large and positive (i.e. the
limit $Q^2\rightarrow \infty$, $\xi$ large but fixed); 2) $t/\xi\rightarrow 0$ (or the Regge limit: $x\rightarrow 0$
and $Q^2$ large but fixed); 3) $t/\xi\rightarrow -\infty$ (or the photoproduction limit: $Q^2\rightarrow 0$ and
$\xi$ large but fixed). In case 1) at $M_0$ the derivative must be large and negative, thus $M_0$ is small (see
fig.~4). One finds:
\beq
M_0=\sqrt{\frac{n_c\as\xi}{\pi t}},~~~~~~~G(\xi,t)\propto\exp{[M_0t+\as\chi_0(M_0)\xi]}\approx 
\exp{\sqrt{\frac{4n_c\as t \xi}{\pi}}}\label{1)}
\eeq
Since we know that $\as t$ at fixed $\as$ corresponds to $\eta/b$ for running $\as$, we see that the double scaling
result, eq.(\ref{res}), is reproduced. In case 2), $t/\xi\rightarrow 0$ and the derivative must vanish at $M_0$, or
$M_0=1/2$. The asymptotic behaviour is fixed by  $\chi_0(1/2)$ (see eq.(\ref{chi02})) and we have:
\beq
G(\xi,t)\propto\exp{[\as\chi_0(1/2)\xi]}\approx 
x^{-\lambda_0},~~~~~~\lambda_0=\as\frac{n_c}{\pi}4\ln{2}\label{2)}
\eeq
This is the hard Pomeron prediction (as opposed to the soft Pomeron, or $\lambda_0\approx 0$),which is well known but
of dubious physical relevance, as we shall see in the following. In case 3) $t/\xi$ is large and negative, so the
derivative must be large and positive, which leads to $M_0\approx 1$ where $\chi_0(M)\approx n_c/[\pi (1-M)]$. A
simple calculation leads to:
\beq
M_0=1-\sqrt{\frac{n_c \as \xi}{\pi |t|}},~~~~~~G(\xi,t)\propto Q^2 \exp{\sqrt{\frac{4n_c\xi \as |t|}{\pi}}}\label{3)}
\eeq
This result is qualitatively encouraging, because it leads to the correct linear vanishing of $G(\xi,t)$ in $Q^2$ in
the photoproduction limit~\cite{afp}. But, quantitatively we cannot trust this result, because non perturbative effects must be
important at $Q^2\approx 0$. Actually this confirms that the BFKL function must contain non perturbative effects near
$M=1$ otherwise photoproduction would also be perturbative.

\section{ The Duality Relations}
\noindent
In the region where
$Q^2$ and
$1/x$ are both large the
$t$ and
$\xi$ evolution equations, \ie~eqs.(\ref{tevol},\ref{xevol}),   are simultaneously valid, and their mutual
consistency requires the validity of the ``duality" relation~\cite{jar,afp}:
\beq
\chi(\gamma(N,\as),\as)=N,\label{dual}
\eeq and its inverse
\beq
\gamma(\chi(M,\as),\as)=M.\label{revdual}
\eeq
In order to derive the duality relations we start from the $x$ evolution equation, eq.(\ref{xevol}), and take the
$x$ Mellin transform of both sides:
\beq
\int^\infty_0~d\xi~e^{-N\xi}~\frac{d}{d\xi}~G(\xi,M)=\chi(M,\as)~G(N,M)\label{dMT}
\eeq
Integration by parts leads to:
\beq
[e^{-N\xi}~G(\xi,M)]^\infty_0+NG(N,M)=\chi(M,\as)~G(N,M)\label{dMT1} 
\eeq
The square bracket on the l.h.s is determined by $\xi\sim 0$ and provides an N independent boundary condition, which
we denote by $-H_0(M)$ (recall that the $x$ evolution equation is only valid at large $\xi$). Solving for $G(N,M)$, we
find:
\beq
G(N,M)=\frac{H_0(M)}{N-\chi(M,\as)}\label{dMT2}
\eeq
The position of the singularity in $N$ fixes the large $t$ behaviour of $G(N,t)$:
\beq
G(N,t)=\int_{c-i\infty}^{c+i\infty}\!\frac{dM}{2\pi i}~\exp{Mt}~G(N,M)
\toinf{t}
\chi(M_p,\as)=N\label{dMTx}
\eeq
while the contribution of additional singularities further on the left of $M_p$ is suppressed by powers of $Q^2$. Then, at
fixed
$\as$, this must coincide with
\beq
G(N,t)\toinf{t}
e^{\gamma(N,\as)t}\label{gaga}
\eeq
where $\gamma$ is the anomalous dimension function at all order in $\as$. Thus $M_p=\gamma(N,\as)$ and the duality relation
eq.(\ref{dual}), and its inverse, eq.(\ref{revdual}), are obtained. These relations are true for the complete leading
twist contribution in the domain where the
$t$ and
$\xi$ evolution equations, \ie~eqs.(\ref{tevol},\ref{xevol}),   are simultaneously valid.
\begin{figure}
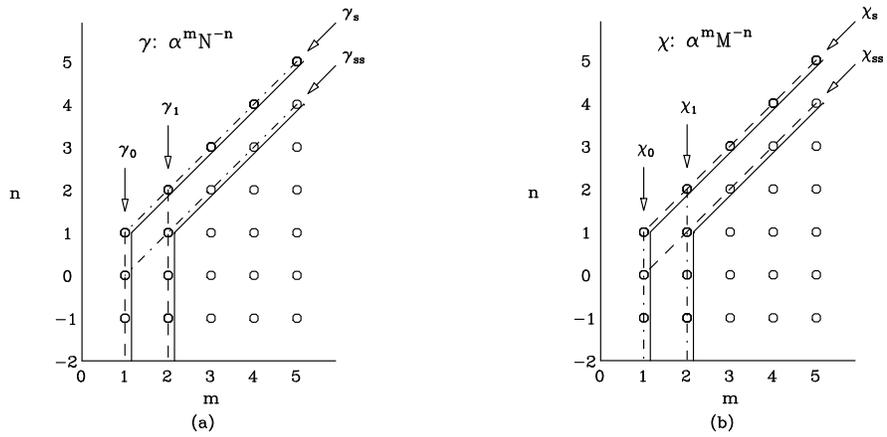

\begin{center}
\mbox{\hbox{
 \epsfig{file=fig1ax.ps,width=0.4\textwidth}}
\hskip 2truecm\hbox{\epsfig{file=fig1bx.ps, width=0.4\textwidth}}}
\caption{\baselineskip 10pt Graphical representation of different  expansions of (a) $\gamma$  and (b) $\chi$  in
powers of 
$\as$ and $1/N$ (a) and $1/M$ (b) to order $m$ and $n$ respectively,  and the different relations between these
expansions.  Vertical lines correspond to terms of the same fixed order in $\as$: for example the one loop anomalous
dimension $\gamma_{0}$ contains terms with $m=1$, $n=1,0,-1,-2,\ldots$.  Diagonal lines correspond to terms of the
same order in 
$\as$ at fixed $\as\over N$ (a) or $\as\over M$ (b): for  example $\gamma_s(\as/N)$  contains terms with
$m=n=1,2,3,\ldots$.  The sum of terms in a vertical line of the $\gamma$ plot is related by duality 
eqs.~(\ref{dual},\ref{revdual}) to the sum of terms  in a diagonal line in the $\chi$  plot and conversely (marked by
the same line style). The solid lines denote terms of the same order in the ``envelope'' or ``double leading''
expansion discussed in the text. }{ }
\end{center}
\vskip -1truecm
\end{figure}

Using eq.~(\ref{dual}), knowledge of the expansion  eq.~(\ref{chiexp}) of $\chi(M,\as)$  to LO and NLO in $\as$
at fixed $M$  determines the coefficients of the expansion of 
$\gamma(N,\as)$ in powers of $\as$ at fixed $\smallfrac{\as}{N}$:
\beq
\gamma(N,\as)=\gamma_s(\smallfrac{\as}{N}) +\as\gamma_{ss}(\smallfrac{\as}{N})+\ldots,\label{sxexp} 
\eeq where $\gamma_s$ and $\gamma_{ss}$ contain respectively sums of  all the leading and subleading singularities of
$\gamma$  (see fig.~3). To derive the expressions of $\gamma_s$ and $\gamma_{ss}$ we start from eq.(\ref{dual})
written in the form:
\beq 
\as \chi_0[\gamma_s(\smallfrac{\as}{N}) +\as\gamma_{ss}(\smallfrac{\as}{N})+...]+\as ^2
\chi_1[\gamma_s(\smallfrac{\as}{N}) +...]+....=N\label{sxexp1}
\eeq
We then expand in $\as$ at $\as/N$ fixed:
\beq
\chi_0(\gamma_s(\smallfrac{\as}{N}))+\as \chi'_0(\gamma_s(\smallfrac{\as}{N}))~\gamma_{ss}(\smallfrac{\as}{N})+
\as \chi_1(\gamma_s(\smallfrac{\as}{N}))+....=\frac{N}{\as}\label{expexp}
\eeq
{}From this we find the relations:
\bea
\chi_{0}(\gamma_{s}(\smallfrac{\as}{N}))&=&{N\over\as},\label{dual0}\\
\gamma_{ss}(\smallfrac{\as}{N})&=& -\frac{\chi_{1}(\gamma_{s}(\smallfrac{\as}{N}))}
{\chi'_{0}(\gamma_{s}(\smallfrac{\as}{N}))}.\label{dual1}
\eea This corresponds to an expansion of the splitting function  in logarithms of $x$: if for example we write
\beq
\gamma_{s}(\frac{\as}{N})=
\sum_{k=1}^{\infty}g_k^{(s)}\left(\frac{\as}{N}\right)^k
\label{ga0}
\eeq  then the associated
splitting function is given by
\beq P_s(\as\xi)\equiv  \int_{c-i\infty}^{c+i\infty}\!\frac{dN}{2\pi i\as}\, e^{N\xi}\,\gamma_s(\smallfrac{\as}{N})=
\sum_{k=1}^{\infty}\frac{g_k^{(s)}}{(k-1)!}(\as\xi)^{(k-1)},\label{Pa}
\eeq and similarly for the subleading singularities $P_{ss}(\as\xi)$, etc. From eq.(\ref{chi01}) it follows that
\beq
g_1^{(s)}=n_c/\pi,\>g_2^{(s)}=g_3^{(s)}=0\>, g_4^{(s)}=2 \zeta(3) n_c/ \pi,\dots\label{num}
\eeq
We see that the logaritmic terms in $P_s$ only start at order $\as^3$ due to the vanishing of $g_2^{(s)}$ and
$g_3^{(s)}$. It is also important to observe that the coefficients of the $P_s$ expansion are factorially suppressed
with respect to those of $\gamma_s$. This implies that while the convergence radius of the $\gamma_s$ expansion is
finite, that of $P_s$ is infinite. 

At this point it is interesting to make a digression and compare the small $x$ behaviour of the singlet structure
functions in the spacelike region with that of singlet fragmentation functions in the timelike region. In the leading
log approximation the timelike splitting functions, relevant for the evolution of fragmentation functions,  are the
same as the spacelike splitting functions. Therefore the solution to the one loop evolution equation in the
small $x$ region is valid both for parton distributions in the spacelike region and for fragmentation functions in the
timelike region. But if this were the correct asymptotic result at $x\rightarrow 0$ for the singlet framentation
functions it would imply that the average multiplicity (the $N=0$ moment) is singular. Indeed  it is in fact known
\cite{bass} that, in the timelike
region, the behaviour of the anomalous dimension near $N=0$ is actually modified by higher order terms according
to:
\beq
\gamma(N,\as)=\frac{\as n_c}{\pi N}-2(\frac{\as n_c}{\pi})^2\frac{1}{N^3}+......=
\frac{1}{4}[-N+\sqrt{N^2+\frac{8\as n_c}{\pi}}]\label{tim}
\eeq
In $x$ space this is equivalent to the occurrence of terms of order $\as[\as \log^2{1/x}]^{n-1}$ in nth order
perturbation theory in $\as$. Thus there are two powers of log for each $\as$ in the timelike region instead
than one. These terms modify the behaviour of
$\gamma(N,\as)$ near
$N=0$ into a non singular one. This corresponds to the well known behaviour of the average multiplicity in gluon jets
given by:
\beq
<n_g>\propto~\exp{\int_{\as} ^{\as(t)}~d\as~\frac{\gamma(0,\as)}{\beta(as)}}\approx
\exp{\sqrt{\frac{2 n_c \log{Q^2/\Lambda^2}}{\pi b \log{\mu^2/\Lambda^2}}}}\label{molt}
\eeq
which is well supported by experiment. How is it possible that there is such a difference between the spacelike and
timelike regions? That the fragmentation functions are more singular than the structure functions is due to the
fact that in leptoproduction the final state is totally inclusive. Instead, when we compute the fragmentation of a gluon
into a gluon, we fix the momentum of the observed soft gluon and so its associated infrared singularity cannot be canceled
against the corresponding virtual contribution and appears in the result as an extra power of $\xi$. 

Going back to the duality relations, the inverse duality eq.~(\ref{revdual}) relates the fixed order expansion
eq.~(\ref{gamexp}) of
$\gamma(N,\as)$ to  an expansion of $\chi(M,\as)$ in powers of $\as$ with 
$\smallfrac{\as}{M}$ fixed: if 
\beq
\chi(M,\as) =\chi_s(\smallfrac{\as}{M})+\as\chi_{ss}(\smallfrac{\as}{M}) +\ldots,\label{clqexp} 
\eeq where now $\chi_s(\smallfrac{\as}{M})$ and 
$\chi_{ss}(\smallfrac{\as}{M})$ contain the leading  and subleading singularities respectively of $\chi(M,\as)$, then
\bea
\gamma_{0}(\chi_{s}(\smallfrac{\as}{M}))&=&\frac{M}{\as},
\label{revdual0}\\
\chi_{ss}(\smallfrac{\as}{M})&=& -\frac{\gamma_{1}(\chi_{s}(\smallfrac{\as}{M}))}
{\gamma'_{0}(\chi_{s}(\smallfrac{\as}{M}))}.
\label{revdual1}
\eea
We now discuss the physical implications of these formal results for the singlet splitting function at small $x$. 

\section{Improving the BFKL Expansion}
\noindent
In principle, since $\chi_0$ and $\chi_1$ are known, they can be  used to construct an
improvement of the splitting function which  includes a summation of leading and subleading logarithms of $x$. 
However, as is now well known,  the calculation~\cite{fl,cc,dd} of $\chi_{1}$  has shown that this procedure is
confronted with serious problems. The fixed order expansion eq.~(\ref{chiexp}) is very badly behaved: at  relevant
values of $\as$ the NLO term completely overwhelms the LO term.  In particular, near $M=0$, the behaviour is
unstable, with $\chi_0\sim 1/M$,
$\chi_1\sim -1/M^2$.  Also, the value of $\chi$ near the minimum is subject to a large negative NLO correction, 
which turns the minimum into a maximum and can even reverse the sign of $\chi$ at the minimum:  in $\MS$ for $n_f=3$
or $4$, we approximately have
\beq
\chi(\smallfrac{1}{2},\as)\approx 2.65\as (1-6.2\as +...)\label{numb}
\eeq
 Finally, if one
considers the resulting $\gamma_s$ and $\gamma_{ss}$ or their Mellin transforms 
$P_s(x)$ and $P_{ss}(x)$ one finds that the NLO terms become much larger than the LO terms and negative in the region
of relevance for the HERA data~\cite{flph}.  We now discuss our proposals to deal with all these problems.

Our first observation is that a much more stable expansion for $\chi(M)$  can be obtained if we make appropriate use
of the additional information  which is contained in the one and two loop anomalous dimensions 
$\gamma_{0}$ and $\gamma_{1}$. Instead of trying to improve the fixed  order expansion eq.~(\ref{gamexp}) of $\gamma$
by all order summation  of singularities deduced from the fixed order expansion eq.~(\ref{chiexp})  of $\chi$, we
attempt the converse: we improve 
$\chi_{0}(M)$ by adding to it the all order summation of  singularities $\chi_{s}$ eq.~(\ref{revdual0}) deduced  from
$\gamma_0$, $\chi_{1}(M)$ by adding to it $\chi_{ss}$ deduced  from $\gamma_1$  eq.~(\ref{revdual1}), and so on. It
can then be seen that the instability at $M=0$ of the usual fixed order expansion  of $\chi$ was inevitable: momentum
conservation for the  anomalous dimension, eq.~(\ref{mcons}), implies, given the  duality relation, that the value of
$\chi(M)$ at $M=0$ is fixed  to unity, since from eq.~(\ref{dual}) we see that at $N=1$
\beq
\chi(\gamma(1,\as),\as)=\chi(0,\as)=1.\label{dualmcons}
\eeq It follows that the fixed order expansion of $\chi$ must be poorly behaved near $M=0$: a simple  model of this
behaviour is to think of replacing $\as/M$ with 
$\as/(M+\as)=\as/M - \as^2/M^2+...$ in order to  satisfy the momentum
conservation constraint.  

We thus propose a
reorganization of the expansion of $\chi$ into  a ``double leading'' (DL) expansion, organized in terms of
``envelopes''  of the contributions summarized in fig.~3b: each order contains a ``vertical'' sequence of terms of
fixed order in $\as$, supplemented by a ``diagonal''  resummation of singular terms of the same order in $\as$ if
$\as/M$ is  considered fixed. To NLO the new expansion is thus
\bea &\chi(M,\as)&=\left[\as\chi_{0}(M) +\chi_{s}\left({\smallfrac{\as}{M}}\right)-
\smallfrac{n_c\as}{\pi M}\right]\nonumber\\ &&\qquad +\as\left[\as\chi_{1}(M)
+\chi_{ss}\left({\smallfrac{\as}{M}}\right)-\as\left(\smallfrac{f_2}{M^2}+
\smallfrac{f_1}{M}\right)-f_0\right]+\cdots\label{cdl}
\eea where the LO and NLO terms are contained in the respective square brackets.  Thus the LO term contains three
contributions: $\chi_0(M)$ is the  leading BFKL function~eq.~(\ref{chiexp}), $\chi_{s}(\as/M)$  eq.~(\ref{revdual0})
are resummed leading singularities  deduced from the one loop anomalous dimension, and
$n_c\as/(\pi M)$ is subtracted  to avoid double counting. At LO the momentum conservation 
constraint~eq.~(\ref{dualmcons}) is satisfied exactly  because $\gamma_{0}(1)=0$ and
$[\chi_0(M)-\smallfrac{n_c}{\pi M}]\sim M^2$ near $M=0$ (see eq.(\ref{chi01})). At NLO there are again three types of
contributions: 
$\chi_1(M)$ from  the NLO fixed order calculation (eq.~(\ref{chionedef})), the  resummed subleading singularities
$\chi_{ss}(\as/M)$ deduced  from the two loop anomalous dimension,  and three double counting terms, $f_0=0$, $f_1= 
-n_f(13+10n_c^2)/(36\pi^2n_c^3) $ and
$f_2=n_c^2(11+ 2 n_f/n_c^3)/(12\pi^2)$ (corresponding to those terms with
$(m,n)=(1,0),(2,1),(2,2)$ respectively in fig.~3b).  Note that at the next-to-leading level the momentum conservation 
constraint is not exactly satisfied because the constant  contribution to $\chi_1$ does not vanish in $\MS$, even
though  it is numerically very small (see fig.~4). It could be made exactly zero by a refinement of the double 
counting subtraction but we leave further discussion of this point for  later.

\begin{figure}[t!]
\begin{center}
    \mbox{
      \epsfig{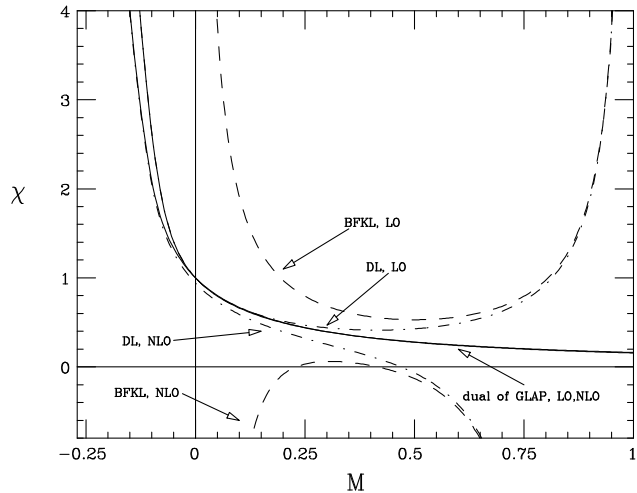}
      }\caption{\baselineskip 10pt 
      Plots of different approximations to the $\chi$ function
      discussed in the text: the BFKL leading and next-to-leading  order functions eq.~(\ref{chiexp}), $\as\chi_0$
and $\as\chi_0+\as^2\chi_1$ (dashed); the LO and NLO dual $\as \chi_{s}$ and 
$\as \chi_{s}+ \as^2 \chi_{ss}$  of the one and two loop anomalous dimensions (solid), and the double-leading
functions at LO and NLO defined in   eq.~(\ref{cdl}) (dotdashed). Note that the double-leading curves coincide with 
the resummed ones at small $M$, and with the fixed order ones at large $M$. }{}
\end{center}
\end{figure}

Plots of the various LO and NLO approximations to $\chi$ are  shown in fig.~4. In this and other plots in this paper
we take
$\as =0.2$, which is a typical value in the HERA region, and the number of active flavours $n_f=4$.  We see that, as
discussed above, the usual fixed order expansion eq.~(\ref{chiexp}) in terms of $\chi_0$  and $\chi_1$ is  very
unstable. However, the new  expansion eq.~(\ref{cdl}) is stable up to $M\lappeq 0.3-0.4$.  Furthermore, in this
region, $\chi$ evaluated in the double leading  expansion (\ref{cdl}) is very close to the resummations of leading
and  subleading singularities eq.~(\ref{clqexp}) obtained by duality  eq.~(\ref{revdual0},\ref{revdual1}) from the
one and two loop  anomalous dimensions. This shows that in this region the dominant  contribution to $\chi$, and thus
to $\gamma$ , comes from the  resummation of logarithms of $Q^2/\mu^2$ with $Q^2\gg \mu^2$.

Beyond $M\sim 0.4$, the size of the  contributions from collinear singular and nonsingular terms becomes  comparable
(after all here $Q^2\sim\mu^2$), but the calculation of the latter  (from the fixed expansion eq.~(\ref{chiexp})) has
become unstable due to  the influence of the singularities at $M=1$. No complete  and reliable description of $\chi$
seems possible without some sort of  stabilization of these singularities. However, since they correspond  to
infrared singularities of the BFKL kernel (specifically logarithms of 
$Q^2/\mu^2$ with $Q^2\ll \mu^2$) this  would necessarily  be model dependent. In particular, such a stabilization
cannot be easily deduced from the resummation of the $M=0$ singularities: the original symmetry of the gluon--gluon
amplitude at large $s$ is spoiled by running coupling effects and by unknown effects from the coefficient function
through which it is related  to the deep-inelastic structure functions, in a way which  is very difficult to control
near the photoproduction limit 
$M\to 1$. We thus prefer not to enter into this problem: rather we   will discuss later a practical procedure to
bypass it.

The results summarized in fig.~4 clearly illustrate the superiority  of the new double leading expansion of $\chi$
over the fixed order  expansion, and already indicate that the complete 
$\chi$ function could after all lead to only small departures from  ordinary two loop evolution.

\section{Improving the Splitting Function Expansion}
\noindent
Having constructed a more satisfactory expansion~eq.~(\ref{cdl})  of the kernel $\chi$,
we now derive from it an improved form of the anomalous dimension $\gamma$ to be used in the
evolution~eq.~(\ref{tevol}),  in order to achieve a more complete description  of scaling violations valid both at
large and small $x$. In principle,  this can be done by using the duality relation eq.~(\ref{dual}),  which simply
gives the function $\gamma$ as the inverse of the  function $\chi$. However, in order to derive an analytic 
expression for $\gamma(N,\as)$ which also allows us to clarify the relation to previous attempts we start from the
naive double-leading expansion of 
$\gamma$~\cite{summ} in which  terms are organized into ``envelopes'' of the contributions  summarized in fig.~3a in
an analogous way to the double leading  expansion (\ref{cdl}) of $\chi$: 
\bea &\gamma(N,\as)&=\left[\as\gamma_{0}(N) +\gamma_{s}\left(\smallfrac{\as}{N}\right)-
\smallfrac{n_c\as}{\pi N}\right]\nonumber\\ &&\qquad +\as\left[\as\gamma_{1}(N)
+\gamma_{ss}\left(\smallfrac{\as}{N}\right) -\as\left(\smallfrac{e_2}{N^2}+
\smallfrac{e_1}{N}\right)-e_0\right]+\cdots,\label{gdl}
\eea where now $e_2=g_2^{(s)}=0$, $e_1=g_1^{(ss)}= n_f n_c (5+13/(2n_c^2))/(18\pi^2)$  and
$e_0=-(\smallfrac{11}{2}n_c^3+ n_f)/(6\pi n_c^2)$. In this equation, the leading and subleading singularities
$\gamma_{s}$ and
$\gamma_{ss}$ are obtained using duality eq.~(\ref{dual})  from $\chi_0$ and $\chi_1$, and summed up to give
expressions which  are exact at NLLx. These are then added to the usual one and two loop  contributions, and the
subtractions take care of the double counting  of singular terms. 

It can be shown that the dual of  the double leading expansion of $\chi$ eq.~(\ref{cdl}) coincides  with this double
leading expansion of $\gamma$  eq.~(\ref{gdl})  order by order in perturbation theory, up to terms which are higher
order in the sense of the double leading expansions.  However, it is clear that these additional subleading terms
must be numerically important. Indeed, it is well know that at small $N$ the anomalous dimension in the small-$x$
expansion eq.~(\ref{sxexp}) is completely dominated by $\gamma_{ss}(\as/N)$ which grows very  large and negative,
leading to completely unphysical results in the  HERA region~\cite{flph}. It is clear that this perturbative 
instability will also be a problem in the  double leading expansion eq.~(\ref{gdl}). On the other hand,  we know from
fig.~4 that the exact dual of $\chi$ in double  leading expansion is stable, and  not too far from the usual two loop
result. The  origin of this instability problem, and a suitable reorganization of the  perturbative expansion which
allows the resummation of the  dominant part of the subleading terms have been discussed  in ref.~\cite{sxap}. After
this resummation, the resulting  expression for $\gamma$ in double leading expansion  will be very close to the exact
dual of the corresponding expansion  of $\chi$.

To understand this point we consider the asymptotic behaviour of $P_s$ and $P_{ss}$ at large $\xi$. Starting from
the definition of the splitting function as the inverse Mellin transform of the anomalous dimension $\gamma$ as
given in eq.(\ref{Pa}), we have:
\beq
P_s(\as\xi)\equiv  \int_{c-i\infty}^{c+i\infty}\!\frac{dN}{2\pi i\as}\, e^{N\xi}\,\gamma_s(\smallfrac{\as}{N})=
-\int_{c-i\infty}^{c+i\infty}\!\frac{d\gamma_s}{2\pi i}\,e^{\as \chi_0(\gamma_s)\xi}\,\gamma_s\chi'_0(\gamma_s)
\label{psasy}
\eeq
where we have used eq.(\ref{dual0}) to change the integration variable according to 
$dN=\as \chi'_0(\gamma_s) d\gamma_s$. Similarly for $P_{ss}$, by using the form of $\gamma_{ss}$ given in
eq.(\ref{dual1}), we obtain: 
\beq
P_{ss}(\as\xi)\equiv  \int_{c-i\infty}^{c+i\infty}\!\frac{dN}{2\pi i\as}\, e^{N\xi}\,\gamma_{ss}(\smallfrac{\as}{N})=
+\int_{c-i\infty}^{c+i\infty}\!\frac{d\gamma_s}{2\pi i}\,e^{\as \chi_0(\gamma_s)\xi}\,\chi_1(\gamma_s)
\label{pssasy}
\eeq
The asymptotic behaviour of $P_s$ and $P_{ss}$ at large $\xi$ can now be obtained by applying the saddle point
method. Remembering that the only real minimum of $\chi_0(\gamma_s)$ in the range $0<\gamma_s<1$ is at
$\gamma_s=\smallfrac{1}{2}$, and expanding around the minimum according to
\beq
\gamma_s\chi'_0(\gamma_s)=\smallfrac{1}{2}\chi^{''}_0(\smallfrac{1}{2})(\gamma_s-\smallfrac{1}{2})~+~
\chi^{''}_0(\smallfrac{1}{2})~(\gamma_s-\smallfrac{1}{2})^2+....;~~~~~\chi_1(\gamma_s)=
\chi_1(\smallfrac{1}{2})+....\label{saddle}
\eeq
one finds (see the formulae on the saddle point method in Appendix):
\beq
P_s(\as \xi)\toinf{\xi}
e^{\as \chi_0(\smallfrac{1}{2})
\xi}=x^{-\lambda_0};~~~~~\frac{P_{ss}}{P_s}\toinf{\xi}
\as
\chi_1(\smallfrac{1}{2})\xi.\label{bafo}
\eeq
We see that at sufficiently small values of $x$ $P_{ss}$ overwhelms $P_s$ causing the perturbative expansion to become
unstable. Actually this occurs at not so small values of $x$ because $\chi_1/\chi_0$ is so large.

On the basis of these results the procedure of ref.~\cite{sxap} can be interpreted in a simple way whenever the
all-order ``true'' function
$\chi(M,\as)$  possesses a minimum at a real value of $M$, $M_{min}$,  with $0<M_{min}<1$ (although the final result
for the anomalous dimension will retain its validity even in the absence  of such minimum). Using $\lambda$ to denote
this minimum value of $\chi$,
\beq
\lambda\equiv\chi(M_{min},\as)=\lambda_0+\Delta \lambda, 
\qquad \lambda_0\equiv\as \chi_0(\half)= \smallfrac{4n_c}{\pi} \as \ln 2. 
\label{lamdef}
\eeq The instability turns out to be due to the fact that higher order  contributions to $\gamma$ must change the
asymptotic small $x$ behaviour from $x^{-\lambda_0}$ to $x^{-\lambda}~=~x^{-\lambda_0} e^{\Delta \lambda
\xi}~\approx ~x^{-\lambda_0}[1+\Delta \lambda \xi+....]$, with $\Delta\lambda=\as^2\chi_1(\half)+\cdots$.Having
understood the source of the problem we can now cure it by a suitable resummation procedure.   In terms of splitting
functions the proposed resummed expansion is simply 
\bea xP(x,\as)&=&\as e^{\xi\Delta\lambda} [P_{s}(\as\xi)+\as \tilde P_{ss}(\as\xi)+\dots]\nonumber\\ &=&\as
e^{\xi\Delta\lambda} [P_{s}(\as\xi)+\as P_{ss}(\as\xi) -\xi\Delta\lambda P_{s}(\as\xi)+\dots].\label{Pimpr}
\eea The expansion is now stable~\cite{sxap},  in the sense that it may be shown that 
$\tilde P_{ss}(\as\xi)/P_{s}(\as\xi)$ remains bounded as 
$\xi\to\infty$: subleading corrections will then be small  provided only that $\alpha_s$ is sufficiently small.
Equivalently, this
procedure consists of absorbing  the value of the correction to the value of
$\chi$ at the minimum into the leading order term in the expansion of $\chi$:
\bea
\chi(M,\as)&=&\as\chi_0(M)+ \as^2\chi_1(M)+\dots\nonumber\\
      &=&(\as\chi_0(M)+\Delta\lambda)+ \as^2\tilde\chi_1(M)+\cdots,
\label{tilsub}
\eea where $\tilde\chi_n(M)\equiv\chi_n(M)-c_n$, with $c_n$ chosen so that
$\tilde\chi_n(M)$ no longer leads to an $O(\as^n)$ shift in the minimum. Since the position $M_{min}$ of the
all-order minimum is not known, one must in practice expand it in powers of $\as$ around the leading order value
$M=\half$, so at higher orders the expressions for the subtraction  constants $c_n$ can become quite complicated
functions of $\chi_i$ and  their derivatives at $M=\half$~\cite{sxap}. However at NLO  we have simply
$c_1=\chi_1(\half)$, so 
$\Delta\lambda=\as^2\chi_1(\half)+\cdots$. 

The corresponding expansion of $\gamma$ in resummed leading and  subleading
singularities can also be obtained from the  duality eqs.(\ref{dual0},\ref{dual1},\dots) by treating
$\chi_0+\Delta\lambda$ as the LO contribution to $\chi$, and the  subsequent terms $\tilde\chi_i$ as perturbative
corrections to it. Of course, since the reorganization eq.~(\ref{tilsub}) amounts  to a reshuffling of perturbative
orders, to any finite order the anomalous dimension obtained in this way will be equal to the  old one up to formally
subleading  corrections.  
 Explicitly, we find in place of the previous expansion in sums of  singularities
eq.~(\ref{sxexp}) the resummed expansion
\beq
\gamma(N,\as)=\gamma_s\left(\smallfrac{\as}{N-\Delta \lambda}\right)+
\as\tilde\gamma_{ss}\left(\smallfrac{\as}{N-\Delta \lambda}\right) +\dots,\label{gamimpr} 
\eeq where 
\beq
\tilde\gamma_{ss}\left(\smallfrac{\as}{N-\Delta \lambda}\right)
\equiv \gamma_{ss}\left(\smallfrac{\as}{N-\Delta \lambda}\right) -{\chi_1(\half)\over\chi'_0\left(\gamma_s
\left(\smallfrac{\as}{N-\Delta \lambda}\right)\right)}.
\eeq 
The shift in the denominators from $N$ to $N-\Delta\lambda$ results from combining the exponential $e^{-N\xi}$ of the
Mellin transform with the exponential $e^{\Delta\lambda\xi}$ factored out in eq.(\ref{Pimpr}). 
 
We can thus replace the unresummed singularities $\gamma_s$  and $\gamma_{ss}$ in eq.~(\ref{gdl})  with the resummed
singularities eq.~(\ref{gamimpr}) to obtain a double leading expansion with stable small $x$ behaviour:
\bea 
\gamma(N,\as)&=& \left[\as\gamma_{0}(N)+
\gamma_{s}(\smallfrac{\as}{N-\Delta \lambda}) -\as\smallfrac{n_c}{\pi N}\right]\nonumber \\ &&
+\as\left[\as\gamma_{1}(N) +\tilde\gamma_{ss}(\smallfrac{\as}{N-\Delta \lambda})
-\as(\smallfrac{e_2}{N^2}+\smallfrac{e_1}{N}) - e_0\right]+\cdots.
\label{til}
\eea Momentum conservation is violated by the resummation   because $\gamma_{s}$ and $\gamma_{ss}$ and the
subtraction terms  do not vanish at $N=1$. It can be restored by  simply adding to the constant $e_0$ a further
series of constant terms beginning at $O(\as^2)$: these are all formally subleading in  the double leading expansion.
This constant shift in $\gamma$ is precisely analogous to the shift made on $\chi$ in eq.~(\ref{lamdef}) which
generated the resummation. 

It is important to recognize that there is inevitably an  ambiguity in the double counting subtraction terms  in
eq.~(\ref{til}). For example, at the leading order  of the double leading expansion instead of subtracting  
$ n_c \as\over\pi N$ we could have subtracted 
$n_c\as\over \pi(N-\Delta\lambda)$, since this differs only by formally subleading terms: $\Delta\lambda=O(\as^2)$, so
\beq {\as\over N}={\as\over N-\Delta \lambda}\left(1-{\Delta \lambda\over N
-\Delta\lambda}+\dots\right).\label{tiltilexp}
\eeq Following the same type of subtraction at NLO, the resummed double leading  anomalous dimension may thus be
written as  
\bea 
\gamma(N,\as)&=& \left[\as\gamma_{0}(N)+
\gamma_{s}(\smallfrac{\as}{N-\Delta \lambda}) -\as\smallfrac{n_c}{\pi (N-\Delta \lambda)}\right]\nonumber \\ &&\quad
+\as\left[\as\gamma_{1}(N) +\tilde\gamma_{ss}(\smallfrac{\as}{N-\Delta \lambda})
+\smallfrac{n_c\Delta\lambda}{\pi(N-\Delta \lambda)^2}\right.\\&&\qquad \left.-\as(\smallfrac{e_2}{(N-\Delta \lambda)^2}
+\smallfrac{e_1}{N-\Delta \lambda}) - e_0\right]+\cdots,\nonumber
\label{tiltil}
\eea 
where the new contribution proportional to $n_c\Delta \lambda$ 
in the NLO subtraction is due
to the fact that to NLO accuracy we cannot simply replace $\alpha_s/N$
with $\alpha_s/(N-\Delta \lambda)$, but we must retain the NLO correction
term in eq.~(\ref{tiltilexp}).  
The characteristic feature of this alternative resummation  is that the
fixed order anomalous dimensions $\gamma_{0}$, $\gamma_{1}$  are preserved in their entirety, including the position
of their  singularities. As with the previous expansion eq.~(\ref{til})  momentum conservation may be imposed by
adding to $e_0$ a  series of terms constant in $N$ and starting at $O(\as^2)$.

This completes our procedure of inclusion of the most important part of the subleading corrections, as we shall see
shortly by a direct comparison of the resummed expansions eq.~(\ref{til}) and eq.~(\ref{tiltil}) with the exact dual
of $\chi$ evaluated according to eq.~(\ref{cdl}). In the sequel we will discuss the phenomenology based on the two 
resummed expansions eq.~(\ref{til}) and eq.~(\ref{tiltil}) on an  equal footing, taking the spread of the results as
an indication of the residual ambiguity due to subleading terms. Although formally the  differences between the two
expansions are subleading, we will find that  in practice they may be quite substantial, because 
$\Delta\lambda$ may be large. 
\begin{figure}[t!]
\begin{center}
    \mbox{
      \epsfig{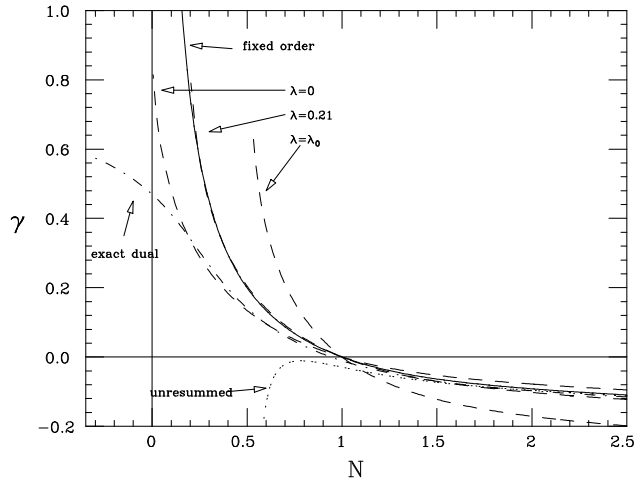}
      }
\caption{\baselineskip 10pt  Comparison of the anomalous dimension $\gamma$ evaluated at NLO in the  resummed
expansion eq.~(\ref{til}) for three different values of $\lambda$   (dashed) with the usual fixed order perturbative
anomalous dimension  (also at NLO) eq.~(\ref{gamexp}) (dotted) and that obtained by  exact duality from $\chi$ at NLO
in the expansion eq.~(\ref{cdl}) as displayed in fig.~4 (solid).  The unresummed $\gamma$ eq.~(\ref{gdl}) is also
shown at NLO. Notice that the $\lambda=0.21$ curve is very close to the two loop anomalous dimension down to the
branch point at $N=\lambda$. }
\vskip -.8truecm {}
\end{center}
\end{figure}

\section{Application to Phenomenology}
\noindent
So far we have constructed resummations of the anomalous  dimension and splitting
function which satisfy the elementary  requirements of perturbative stability and momentum conservation.  This
construction relies necessarily on the value $\lambda$ of $\chi$ near its minimum,  since it is this which determines
the small $x$ behaviour of  successive approximations to the splitting function. In order to obtain a  formulation
that can be of practical use for actual phenomenology,  we will need however to improve the description of $\chi(M)$
in  the ``central region'' near its minimum $M_{min}$, since  as we already observed, we cannot reliably determine
the position and value of the minimum of $\chi$ without a stabilization of the $M=1$ singularity. Indeed, we can see
from fig.~4 that in the central region $\chi$ evaluated in the double  leading expansion is dominated by the
presumably unphysical $M=1$ poles of $\chi$, and at NLO this means that it  actually has no minimum, becoming rapidly
negative. However,  one can use the value $\lambda$ of the true $\chi$ at the  minimum as a useful parameter for an
effective description of the $\chi$ function around $M=1/2$. Indeed, $\Delta \lambda$ as estimated from its
next-to-leading order value $\as^2 \chi_1(1/2)$ turns out to be of the same order as
$\lambda_0$ for plausible values of $\as$, a feature which can be also directly seen from fig.~4. This supports the
idea that $\lambda$ and
$\Delta \lambda$ are not truly perturbative quantities: in general we expect that the overall shift of the minimum
will still be of the order of $\lambda_0$ and negative. It is this order transmutation that makes the impact of the
resummations eq.~(\ref{til},\ref{tiltil}), and  the differences between them, quite substantial. 
After the reinterpretation of $\lambda$ as a parametrization of our ignorance of $\chi$ in the region near
$M=\half$ the resulting $\chi$ function is fixed near $M=0$ by momentum conservation and near the central region by the
value of 
$\lambda$. Thus in the region of interest for deep inelastic scattering it cannot be much different from the true $\chi$
function.

\begin{figure}[t!]
\vskip-.1truecm
\begin{center}
    \mbox{
      \epsfig{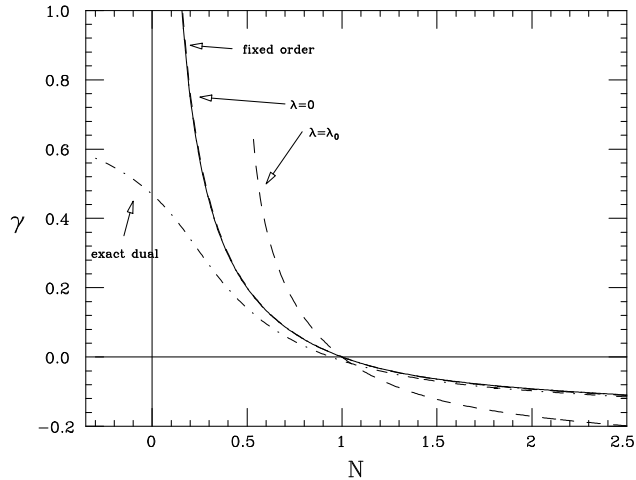}
      }
\caption{
\baselineskip 10pt  Comparison of the anomalous dimension $\gamma$ evaluated at NLO in the  resummed expansion
eq.~(\ref{tiltil}) for two different values of $\lambda$  (dashed) with  the usual fixed order perturbative anomalous
dimension (also at NLO)  eq.~(\ref{gamexp}) (solid) and that obtained by exact duality from  
$\chi$ at NLO as in fig.~4 (dotdash).  Notice that the $\lambda=0$ curve is virtually  indistinguishable from  the
fixed order anomalous dimension  for all values of $N$.}
\vskip -.8truecm {}
\end{center}
\end{figure}

In fig.~5  and fig.~6 we display the results for the resummed anomalous dimensions in the two different expansions
eq.~(\ref{til}) and  eq.~(\ref{tiltil}) respectively, each computed at next-to-leading order.  In both figures we
show for comparison the fixed order anomalous dimension
$\as\gamma_{0}(N)+\as^2\gamma_{1}(N)$  eq.~(\ref{gamexp}). Also for comparison, we show the exact dual  of $\chi$
computed at NLO in the double leading expansion  eq.~(\ref{cdl}), obtained from eq.~(\ref{dual}) by exact numerical
inversion. This curve is thus simply the  inverse of the corresponding curve already shown in fig.~4. 

In fig.~5 we show the anomalous dimension computed at NLO using the  resummation eq.~(\ref{til}), for
$\lambda=\lambda_0$ and $\lambda=0$.  The first value  corresponds to  the LO approximation to $\lambda$, while  the
second value is close to the NLO approximation when $\as$ is in the region
$\as\sim 0.1 - 0.2$.  We might expect the value of $\lambda$ as determined by the actual all-order minimum of $\chi$ 
to lie within this range.    Note that,  in general, the resummed anomalous dimension has a cut starting at
$N=\lambda$, which corresponds to the $x^{-\lambda}$ power rise; for this reason  our plots  stop at this value of
$N$.   The $\lambda=0$ curve, corresponding to the next-to-leading order approximation to $\lambda$, is seen to be
very close to the exact dual of $\chi$ at NLO in the expansion eq.~(\ref{cdl}), as already anticipated.  This is to
be contrasted with the corresponding unresummed anomalous dimension eq.~(\ref{gdl}), which is also  displayed in
fig.~5, and is characterized by the rapid fall at small $N$ discussed already in ref.~\cite{flph,sxap}. This
comparison demonstrates that indeed the perturbative reorganization eliminates this pathological steep decrease. The
resummed curve with $\lambda=0$ and the exact dual of $\chi$ become rather different for small $N\lsim 0.2$. However,
this is precisely the range of $N$ which corresponds to the central region of $M$ where we cannot trust the
next-to-leading order determination of $\chi$.  Finally, we show that we can choose a value of $\lambda\simeq 0.21$
such that the resummed anomalous dimension closely reproduces the two loop result down to the branch point at
$N=\lambda$. This shows that the absence of visible deviations from the usual two loop evolution can be accommodated
by the resummed anomalous dimension. However this is not necessarily the best option phenomenologically: perhaps  the
data could be better fitted by a different value of $\lambda$ if a suitable modification of the input parton
distributions is introduced. It is nevertheless clear that large values of $\lambda$ such as $\lambda
\approx\lambda_0$ can be easily excluded within the framework of this resummation, since they would lead to sizeable
deviations from the standard two loop scaling violations in the medium and large $x$ region.

The splitting functions corresponding to the anomalous dimensions of fig.~5 are displayed in fig.~7. The basic
qualitative features are of course preserved: in particular, the curves with small values of  $\lambda=0$ and
$\lambda=0.21$  are closest to the two loop result. However, on a more quantitative level, it is clear that anomalous
dimensions which coincide in a certain range of $N$,  but differ in other regions (such as very small $N$) may lead
to splitting functions which differ over a considerable region  in $x$. In particular, the 
$\lambda=0.21$ curve displays the predicted $x^{-\lambda}$ growth at sufficiently large $\xi$ ($x\lsim 10^{-4}$). The
dip seen in the figure for intermediate values of $\xi$ is necessary in order to compensate this growth in such a way
that  the moments for moderate values of $N$ remain unchanged. Note that the $x^{-\lambda}$ behaviour of the
splitting function at small $x$ is corrected by logs~\cite{sxap}:  
$P_s\toinf\xi \xi^{-3/2} x^{-\lambda}$. If $\lambda=0$ this logarithmic drop provides the dominant large $\xi$
behaviour which appears in the figure.

If the anomalous dimensions are instead resummed as in eq.~(\ref{tiltil}), the results are as shown in fig.~6, again
for the two very different   values of $\lambda$, $\lambda=0$ and $\lambda=\lambda_0$.  When $\lambda=0$ the resummed
anomalous dimension is now essentially indistinguishable from the two loop result. This is due to the fact that the
simple poles at $N=0$ which are now retained in $\gamma_{0}$ and $\gamma_{1}$    provide the dominant small $N$
behaviour. The branch point at 
$N=\lambda$ in $\gamma_s$ and $\gamma_{ss}$ is then relatively subdominant.  This remains of course true for all
$\lambda\le 0$, and in practice also for small values of $\lambda$ such as
$\lambda\lsim 0.1$. When instead $\lambda=\lambda_0$  the result does not differ appreciably from the resummed 
anomalous dimension shown in fig.~5, since now the dominant small $N$ behaviour is given by the branch point at
$N=\lambda_0$, which is not affected by changes in the double counting prescription. Summarizing, the peculiar
feature of the resummation eq.~(\ref{tiltil}) is that it leads to results which are extremely close to usual two
loops for any value of $\lambda\le 0$, without the need for a fine-tuning of $\lambda$.
\begin{figure}
\begin{center}
    \mbox{
      \epsfig{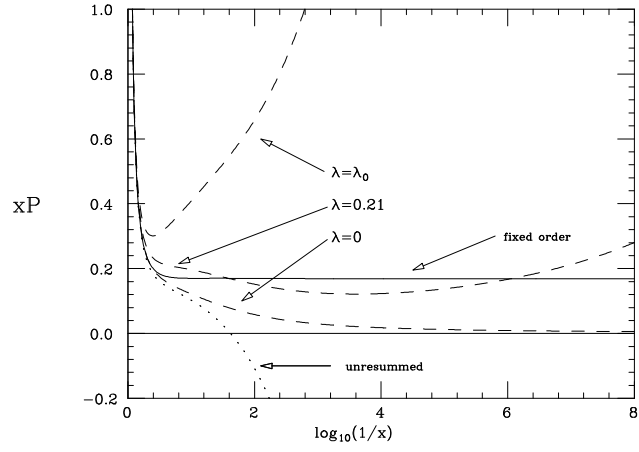}
      }
\caption{
\baselineskip 10pt  The splitting functions corresponding to the anomalous dimensions of fig.~5.}
\vskip -.8truecm {}
\end{center}
\end{figure}

Finally, in fig.~8 we display the splitting functions obtained from the resummed anomalous dimensions of fig.~6. The
$\lambda=\lambda_0$ case is, as expected, very close to the corresponding curve in fig.~7. However the $\lambda=0$
curve is now in  significantly better agreement with the two loop result than any of the resummed splitting functions 
of fig.~7, even that computed with the optimized value
$\lambda=0.21$. Moreover, this agreement now holds in the entire  range of $\xi$. This is due to the fact that the
corresponding  anomalous dimension is now very close to the fixed order one  for all $N>0$, and not only for
$N>\lambda=0.21$. This demonstrates  explicitly that one cannot exclude the possibility that the known small $x$
corrections to splitting functions resum to a result which is essentially indistinguishable from the two-loop one.
This however is only possible if $\lambda\lsim 0$. 
\begin{figure}
\begin{center}
    \mbox{
      \epsfig{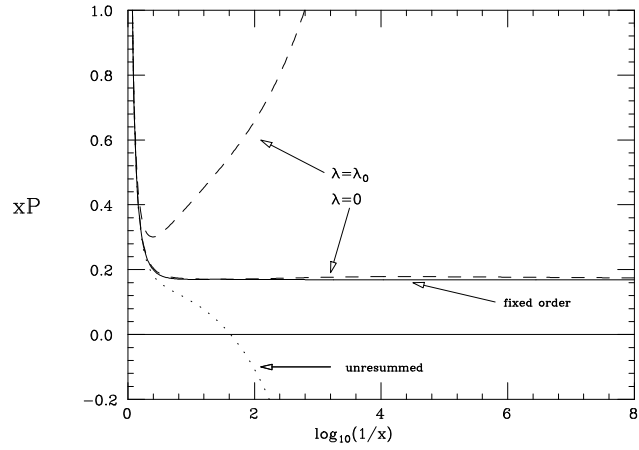}
      }
\caption{
\baselineskip 10pt  The splitting functions corresponding to the anomalous dimensions of fig.~6.}
\vskip -.8truecm {

}
\end{center}
\end{figure}

To summarise, we find that the known success of perturbative evolution, and in particular double asymptotic scaling
at HERA can be accommodated within two distinct possibilities, both of which are compatible with our current
knowledge of anomalous dimensions at small
$x$, and in particular with the inclusion of corrections derived from the BFKL equation to usual perturbative
evolution.  One possibility, embodied by the resummed  anomalous dimension eq.~(\ref{tiltil}) with $\lambda\lsim 0$,
is that double scaling remains a very good approximation to perturbative evolution even if the $x \to 0$ limit is
taken at finite $Q^2$. The other option, corresponding to the resummation eq.~(\ref{til}) with a small value of
$\lambda$, is that double scaling is a  good approximation in a wide region at small $x$, including the HERA region,
but eventually substantial  deviations from it will show up at sufficiently small $x$.  In the latter case, the
best-fit parton distributions might be  significantly differ from those determined at two loops even at the edge of
the HERA kinematic region. Both resummations are however  fully compatible with a smooth matching to Regge theory in 
the low $Q^2$ region~\cite{landshoff}.

\section{Conclusion}
\noindent
In conclusion, we have presented a procedure for the systematic improvement of the
splitting functions at small $x$ which overcomes the difficulties of a straightforward implementation of the BFKL
approach. The basic ingredients of our approach are the following. First, we achieve a stabilization of the
perturbative  expansion of the function $\chi$ near $M=0$ through the resummation of all the LO and NLO collinear
singularities  derived from the known one-- and two--loop anomalous dimensions.  The resulting $\chi$ function is
regular at $M=0$, and in fact, to a good accuracy, satisfies the requirement imposed by momentum conservation via
duality. Then, we acknowledge that without a similar stabilization of the $M=1$ singularity it is not possible to
obtain an reliable determination of $\chi$ in the central region
$M\sim 1/2$. However, we do not have an equally model--independent prescription to achieve this stabilization at
$M=1$.  Nevertheless, the behaviour of $\chi$ in the central region  can be effectively parameterized  in terms of a
single parameter
$\lambda$ which fixes the asymptotic small $x$ behaviour of the singlet parton distribution.  This enables us to
arrive at an analytic expression for the improved splitting function, which is valid both at small and large
$x$ and  is free of perturbative instabilities. 

This formulation  can be directly confronted with the data, which ultimately will provide a determination of
$\lambda$ along with
$\as$ and the input parton densities. The well known agreement of the small $x$ data with the usual $Q^2$ evolution
equations  suggests that the optimal value of $\lambda$ will turn out to be small, and possibly even negative for the
relevant value of
$\alpha_s$.  Such a value  of $\lambda$ is theoretically attractive, because it corresponds to a structure function
whose leading-twist component does not grow as a power of $x$ in the Regge limit: it would  thus be compatible with
unitarity constraints, and with an  extension of the region of applicability of perturbation theory  towards this
limit.

Several alternative approaches to deal with the same problem through  the resummation of various classes of formally
subleading contributions have been recently presented in the literature. Specific proposals  are based on making a
particular choice of the renormalization  scale~\cite{bfklp}, or on a different identification of the large logs
which are resummed by the $\xi$ evolution equation~(\ref{xevol}), either by a function of $\xi$
itself~\cite{schmidt},  or by a function of $Q^2$~\cite{salam,ciaf}, or both~\cite{for}. The main  shortcoming of
these approaches is their model dependence.   For instance, in ref.~\cite{ciaf} the  value of $\lambda$ is
calculated, and $\chi$ is supposedly determined for all 
$0\le M\le 1$. This however requires the introduction of a  symmetrization of $\chi$,  which we  consider to be
strongly model dependent: indeed,  in ref.~\cite{ciaf} it is recognized that their value  of $\lambda$  only signals
the limit of applicability of their computation. We contrast this situation with the approach to resummation
presented here, which makes maximal use of all the available model-independent  information, with a realistic
parameterization of the remaining  uncertainties. We expect further progress to be possible  only on the basis of
either genuinely nonperturbative input, or through a  substantial extension of the standard perturbative domain.

\vspace*{0.9cm}

{\bf Acknowledgement:}  We thank S.~Catani, J.~R.~Cudell, P.~V.~Landshoff, G.~Ridolfi and  G.~Salam for interesting
discussions. G.A. is particularly grateful to Lidia Ferreira for her invitation and the very pleasant and kind
hospitality in Lisbon. This work was supported in part by a PPARC Visiting Fellowship, and  EU TMR  contract FMRX-CT98-0194 (DG
12 - MIHT).

\vspace*{0.9cm}                                                                                           
\noindent
{\bf Appendix: The Saddle Point Method}
\noindent

We want to approximately evaluate the integral  
\beq  I(s)=\int_{c-i\infty}^{c+i\infty}\!\frac{dN}{2\pi i}\!\exp{[s\!f(N)]} g(N)\label{a1}
\eeq  as an expansion in the small parameter $1/s$, where $f(N)$ and $g(N)$ are given, sufficiently well behaved, real
functions.  Assume that there is one (or more) points where $f(N)$ is stationary:
$f'(N_0)=0$. Expanding $f(N)$ and
$g(N)$ around $N_0$ we have
\bea
f(N)=f(N_0)+\half f''(N_0)(N-N_0)^2+......\\ \nonumber
g(N)=f(N_0)+ g'(N_0)~(N-N_0)+ \half g''(N_0)~(N-N_0)^2+.....\label{a2}
\eea
Performing the gaussian integrals one obtains for each $N_0$ a contribution:
\beq
I(s)=\frac{\exp {s f(N_0)}}{\sqrt{2 \pi}}~[\frac{g(N_0)}{\sqrt{s f''(N_0)}}~-~\half
\frac{g''(N_0)}{\sqrt{(s f''(N_0))^3}}+....]\label{a3}
\eeq
which gives the desired expansion in powers of $1/s$.

\end{document}